%% file: Template.tex
\documentclass{article}
\usepackage{spconf,amsmath,graphicx,hyperref}
\usepackage{amssymb,multirow,booktabs,makecell,subcaption,enumitem}
\usepackage[table]{xcolor}
\usepackage{cite}

\title{Toward Elastic Speech Inference: Training-Free Wake-Word Detection from Pretrained ASR}
\name{Hwayeon Kim, Youngwon Choi, Hyeonyu Kim}
\address{MAUM AI Inc., Republic of Korea}

\begin{document}

%
\maketitle

\begingroup
\renewcommand\thefootnote{}
\footnote{This work has been submitted to the IEEE for possible publication. Copyright may be transferred without notice, after which this version may no longer be accessible.}
\addtocounter{footnote}{-1}
\endgroup

\input{Sections/0_abstract}    
\input{Sections/1_introduction}
\input{Sections/3_method}

\input{Sections/4_experiment}
\input{Sections/5_conclusion}
\input{Sections/6_acknowledgement}

\bibliographystyle{IEEEbib}
\bibliography{strings,refs}

\end{document}

%% file: Sections/0_abstract.tex
\begin{abstract}
Recent ASR development has placed growing emphasis on generalization across diverse domains and acoustic conditions. Existing approaches typically adapt pretrained ASR models to front-end functions such as wake-up word (WuW) detection through additional training or task-specific modules. In this work, we explore the use of a shared pretrained ASR backbone for WuW detection without gradient-based fine-tuning and examine whether a compact encoder can be extracted using the PCA-based structured pruning approach of SliceGPT. Experiments with Parakeet-TDT-0.6B-v3 and Moonshine-base show that WuW detection performance remains relatively stable when the encoder channel dimension is reduced by 50\%. These results suggest that task-relevant compact encoders can be derived from pretrained ASR models without fine-tuning.

\end{abstract}


%% file: Sections/1_introduction.tex
\section{Introduction}
\label{sec:intro}

Automatic speech recognition (ASR) models have grown substantially both in model capacity and in training data scale \cite{radford2023robust,seed_asr,sekoyan2025canary,qwen3asr}. These models provide rich speech representations, with increasingly capable pretrained ASR models becoming publicly available. This motivates the question of whether a pretrained ASR model can be leveraged for multiple speech processing functions, such as voice activity detection (VAD), wake-up word (WuW) detection, and ASR, without task-specific fine-tuning.

\begin{figure}[t]
    \centering
    \includegraphics[trim={0 0 0 6.8cm}, clip, width=\columnwidth]{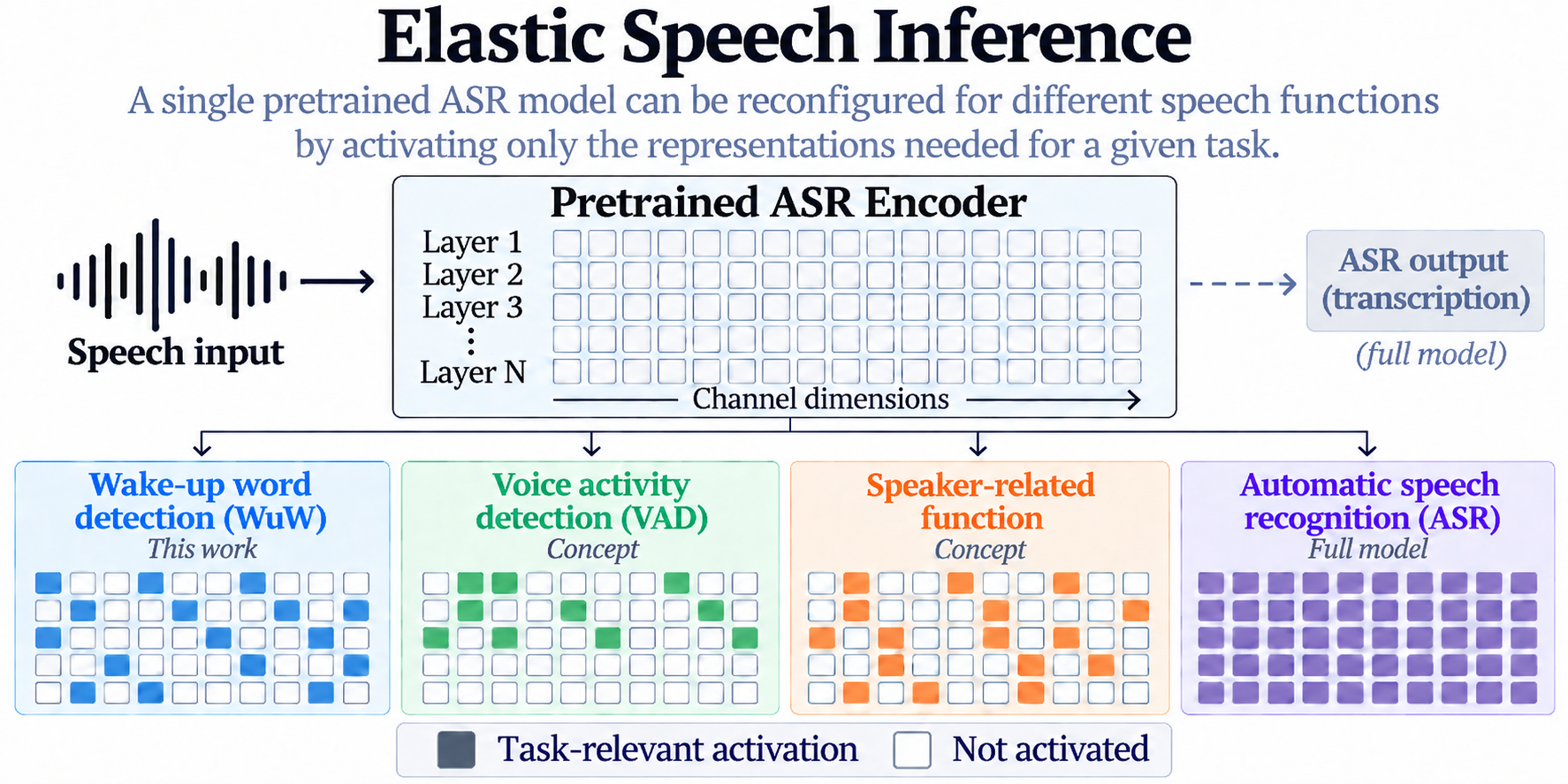}
    \caption{Concept of elastic speech inference. A pretrained ASR encoder is reconfigured by selectively using task-relevant subsets of its internal representations. The activation patterns are schematic and do not correspond to specific channel indices or exact subnetwork configurations.}
    \vspace{-14pt}
    \label{fig:intro}
\end{figure}

Conventional speech systems typically employ separate models for front-end functions, allowing each component to be optimized independently. To better exploit pretrained speech representations for these functions, previous work has adapted pretrained ASR models to downstream tasks, including speaker verification, speech enhancement, and keyword spotting, through fine-tuning or additional modules \cite{cai2024leveraging,zhao2024whisperpmfa,moller2021controlling,eickhoff2023bring,mekonnen2022lowresource,navon2024adakws}. Task-specific subnetworks have also been explored through structured pruning \cite{peng2023structured,genova2025keep,peng2026hybrid}, although these approaches still rely on gradient-based optimization with downstream objectives. More recently, \cite{li2026uaf} integrated several front-end functions, including VAD, turn-taking detection, speaker recognition, and ASR, within a unified model using dedicated heads. However, these approaches also require parameter optimization for individual functions. The extent to which pretrained ASR models can be directly repurposed for front-end tasks without such optimization remains less understood. We refer to this broader direction, where different front-end functions selectively use task-relevant representations from a shared pretrained speech backbone, as \emph{elastic speech inference}. In this setting, the ASR model can remain unchanged while task-relevant subnetworks support each function, avoiding potential degradation of the original ASR capability during downstream adaptation \cite{li2024kwswhisper}. Fig.~\ref{fig:intro} illustrates this concept.

In this work, we focus on WuW detection and investigate whether a detector can be constructed directly from pretrained ASR representations. We refer to this setting as training-free WuW detection, where the pretrained model is used without gradient-based fine-tuning. While query-by-example KWS without keyword-specific model training has been explored using enrollment utterances from a single user \cite{kim2019qbe}, our setting considers speaker-independent detection. We further examine how much the encoder channel dimension can be reduced while preserving WuW performance by adapting SliceGPT's subspace reparameterization, which removes low-energy principal directions without retraining \cite{ashkboos2024slicegpt}, to pretrained ASR encoders. Our experiments show that performance remains relatively stable even with a 50\% reduction in encoder channel dimension, suggesting that only a subset of the original representation capacity may be sufficient for WuW detection.

The main contributions of this work are as follows:

\begin{itemize}[itemsep=0pt, topsep=-2pt]
    \item We demonstrate WuW detection directly from pretrained ASR encoder representations without additional fine-tuning.
    \item We show that PCA-based structured pruning can reduce the encoder channel dimension by 50\% while largely retaining WuW detection performance.
    \item We evaluate the compact encoder across different calibration sources and in a streaming setting, assessing its applicability beyond offline detection.
\end{itemize}

%% file: Sections/3_method.tex
\section{Training-Free Wake-Up Word Detection from Pretrained ASR}

We extract a compact WuW detector from a pretrained ASR encoder without task-specific fine-tuning. The framework combines calibration-based structural compression with enrollment-based similarity matching.

\subsection{Problem Formulation}
Let $f_{\theta}$ denote a pretrained ASR encoder. For an input utterance $x$, the encoder produces
$\mathbf{H}=f_{\theta}(x) \in \mathbb{R}^{T\times d}$
where $T$ and $d$ denote the temporal length and hidden dimension, respectively. Temporal mean pooling yields an utterance-level representation
$\mathbf{z}(x)=P(\mathbf{H}) \in \mathbb{R}^{d}$.

The detection score is calculated as the cosine similarity $s_k(x)$ between $z(x)$ and the enrollment embedding $r_k$. The embedding of the enrollment for the keyword $k$ is defined as

\begin{equation}
\label{eq:prototype}
\mathbf{r}_k
=
\frac{1}{N_e}
\sum_{i=1}^{N_e}
\mathbf{z}\left(x_i^{(k)}\right),
\end{equation}

where $N_e$ is the number of enrollment utterances for each keyword.

The target wake-up word is detected when $s_k(x) \geq \tau_k$, where $\tau_k$ denotes the decision threshold for the keyword $k$. 

\subsection{Orthogonal Subspace-Based Model Compression}

Adapting SliceGPT~\cite{ashkboos2024slicegpt}, we compress
the encoder by rotating hidden representations into
data-dependent orthogonal bases and removing directions
with low activation energy. 

For each selected hidden state location $l$, calibration
activations are collected as
\begin{equation}
\mathbf{X}^{(l)} \in \mathbb{R}^{N_l \times d_l},
\end{equation}
where $N_l$ is the number of collected representations
and $d_l$ is the hidden dimension. We obtain an orthogonal
matrix $\mathbf{Q}^{(l)}$ through uncentered PCA:
\begin{equation}
\mathbf{C}^{(l)}
= {\mathbf{X}^{(l)}}^\top \mathbf{X}^{(l)},
\qquad
\mathbf{Q}^{(l)}
= \left[
\mathbf{q}^{(l)}_1,\ldots,\mathbf{q}^{(l)}_{d_l}
\right],
\end{equation}
where the columns of $\mathbf{Q}^{(l)}$ are the eigenvectors
of $\mathbf{C}^{(l)}$, ordered by decreasing the eigenvalue.

For a hidden-dimension pruning ratio $\gamma$, we retain
\begin{equation}
d'_l = \left\lfloor (1-\gamma)d_l \right\rfloor,
\qquad
\mathbf{Q}^{(l)}_\gamma
= \left[
\mathbf{q}^{(l)}_1,\ldots,\mathbf{q}^{(l)}_{d'_l}
\right].
\end{equation}
Using column-vector notation, the retained coordinates
of a hidden representation are
\begin{equation}
\tilde{\mathbf{h}}^{(l)}
= {\mathbf{Q}^{(l)}_\gamma}^{\top}\mathbf{h}^{(l)}
\in \mathbb{R}^{d'_l}.
\end{equation}

where $\mathbf{h}^{(l)}$ denotes the hidden representation before subspace transformation at location $l$.

Since the representations connected by a residual connection use different subspace bases, a basis mapping is required between them, given by
\begin{equation}
\tilde{\mathbf{R}}^{(l)}
= {\mathbf{Q}^{(l+1)}_\gamma}^{\top}\mathbf{Q}^{(l)}_\gamma.
\end{equation}
This aligns the two reduced subspace coordinates and introduces an additional matrix multiplication during inference. The remaining basis transformations are absorbed into adjacent linear weights.

The resulting encoder operates with reduced hidden dimensions and smaller weight matrices. PCA estimation and weight reparameterization are performed only during compression; no gradient-based fine-tuning is used.

%% file: Sections/4_experiment.tex
\section{Experiments}

\subsection{Experimental Setup}
\subsubsection{Datasets}
We use Google Speech Commands v2 (GSC v2) \cite{warden2018speech} test set for offline wake-up word detection. GSC v2 is a keyword spotting dataset consisting of short English words spoken by multiple speakers, with each utterance lasting approximately 1 s. For each target keyword, its utterances are treated as positive samples, while utterances from all other keywords serve as negatives.

Streaming performance is evaluated on Hey-Snips \cite{coucke2019efficient} test set, a wake-up word dataset containing positive and negative utterances collected from various speakers and recording conditions, with "Hey Snips" as the target phrase. We use this dataset for continuous-stream evaluation.

\subsubsection{Models and Implementation}

We consider two pretrained ASR models: Moonshine-base\cite{jeffries2024moonshine} and Parakeet-TDT\cite{sekoyan2025canary}. These models differ substantially in model scale and ASR architecture, with Moonshine-base using an encoder-decoder formulation and Parakeet-TDT employing a transducer-based architecture. This allows us to examine the proposed approach across pretrained encoders obtained from different ASR architectures and model scales.

The GSC v2 experiments draw 40 enrollment and 512 calibration utterances per keyword from the training split. The enrollment and calibration subsets are disjoint. The Hey-Snips experiments include 40 training-set utterances for enrollment and 501.5 s of LibriSpeech train-clean-100 for calibration, comparable to the duration of the GSC v2 calibration set for each keyword.

\subsubsection{Metrics and Evaluation Settings}

Offline detection performance is measured using equal error rate (EER), false rejection rate (FRR), and false acceptance rate (FAR). EER is computed directly from the test-set FAR--FRR curve. The operating threshold is selected on the validation set to minimize FRR subject to FAR $\leq$ 1\%, and then applied to the test set. Reported results are macro-averaged across all 35 keywords.

For streaming evaluation, we report event-level FRR on the mixed stream and false alarms per hour (FA/h) on the negative-only stream. Audio is processed using a 1.5 s window with a 0.1 s hop. The mixed stream inserts 0.5 s of silence between utterances, while the negative-only stream concatenates negatives without silence. Above-threshold windows are grouped into a single event, with a new event requiring a below-threshold score and at least 1.0 s since the previous detection. A positive utterance is counted as detected if an event occurs after its onset and no later than 0.5 s after its offset.

The validation streams contain 150 positive and 5,000 negative utterances from the validation split. The threshold minimizing FRR under FA/h $\leq$ 0.5 is selected on the validation streams and then applied to the test streams. Compression efficiency is evaluated in terms of encoder parameter count and real-time factor (RTF). RTF is defined as the average per-window processing time divided by the 0.1 s hop duration, with processing time measured on an AMD EPYC 7742 processor using 8 CPU cores. All inference is performed with a batch size of 1.

\begin{figure}[t]
    \centering
    \begin{subfigure}[t]{0.48\columnwidth}
        \centering
        \includegraphics[width=\linewidth]{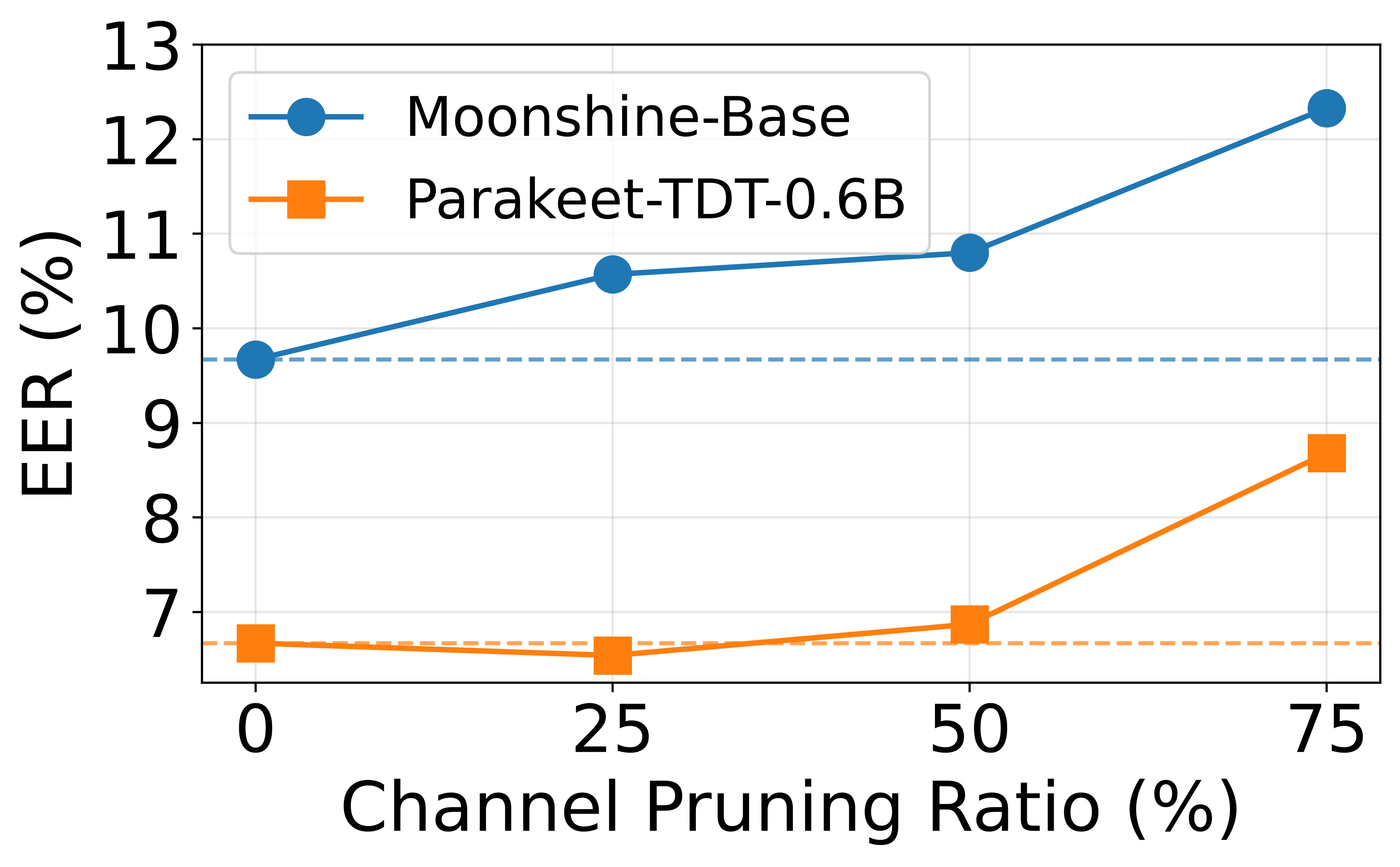}
        \caption{}
        \label{fig:eer}
    \end{subfigure}
    \hfill
    \begin{subfigure}[t]{0.48\columnwidth}
        \centering
        \includegraphics[width=\linewidth]{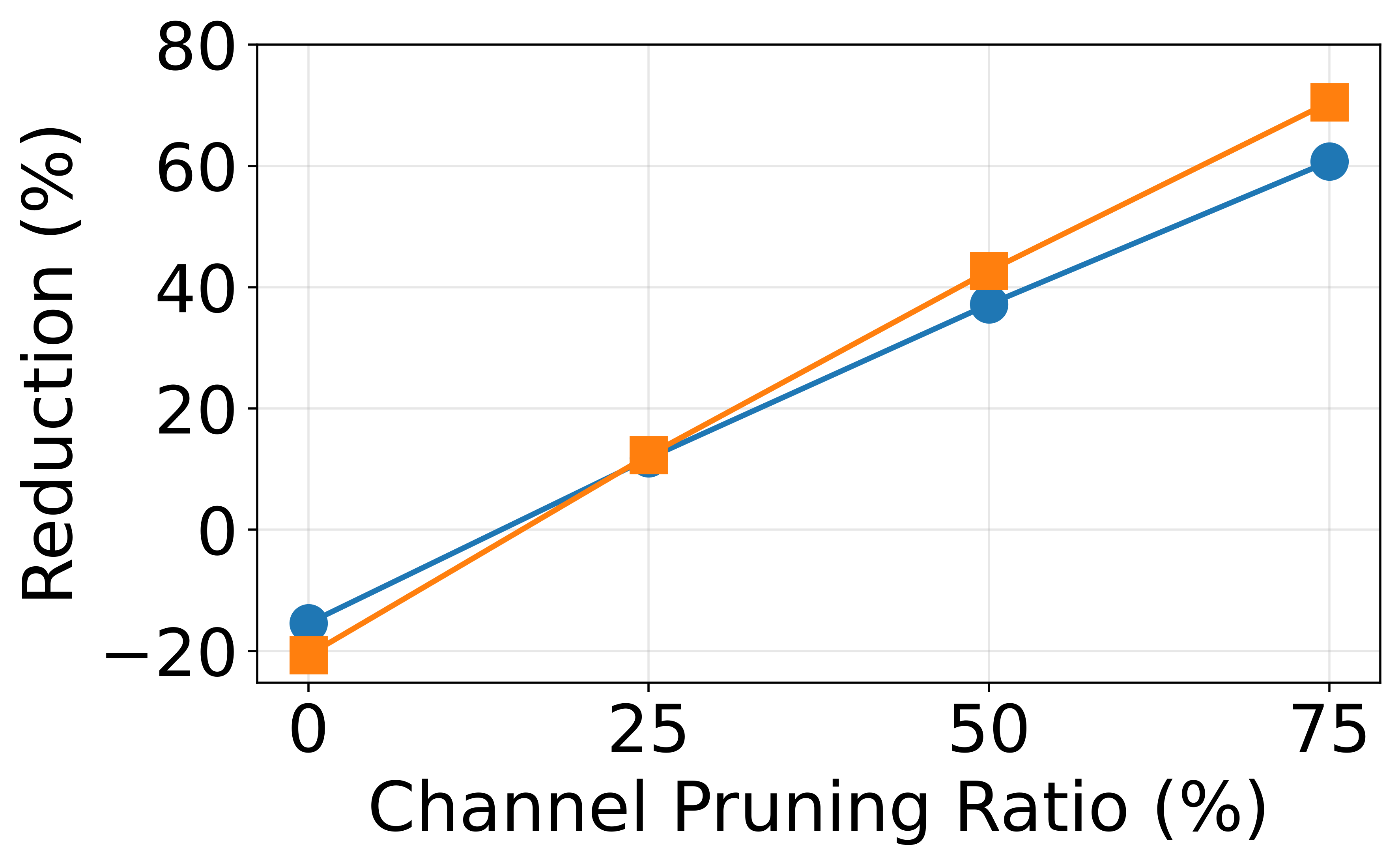}
        \caption{}
        \label{fig:parameter_reduction}
    \end{subfigure}
    \vspace{-8pt}
    \caption{Results on the Google Speech Commands v2 test set across channel pruning ratios: (a) EER and (b) Encoder parameter reduction ratio. Dashed lines in (a) indicate the EERs of the original encoders.}
    \vspace{-14pt}
    \label{fig:pruning}
\end{figure}
\subsection{Compression--Performance Trade-off}

Fig.~\ref{fig:pruning} summarizes the EER and encoder parameter reduction of the two models across different channel pruning ratios. As shown in Fig.~\ref{fig:pruning}(a), at 0\% pruning, the transformed models matched the EER of the original encoders, confirming that the PCA-based transformation itself does not degrade detection performance. Meanwhile, Fig.~\ref{fig:pruning}(b) shows a slight increase in encoder parameter count at 0\% due to additional matrices required at residual connections to align the transformed subspaces. At 50\% channel pruning, both models showed modest increases in EER compared to their original counterparts while reducing the encoder parameter count by approximately 40\%. At the 75\% setting, EER increased more sharply for both models. We therefore selected a 50\% pruning ratio as the operating point for subsequent experiments.

\subsection{Effect of Subspace Selection}

\begin{table}[t]
\centering
\caption{Comparison of PCA-based reduction and random channel selection at 50\% pruning. Mean and std values are computed over 3 random seeds.}
\vspace{-5pt}
\label{tab:pca_vs_random}
\small
\setlength{\tabcolsep}{2.8pt}
\renewcommand{\arraystretch}{1.2}
\begin{tabular}{ccccc}
\hline
Model & Method & EER (\%) $\downarrow$ & FRR (\%) $\downarrow$ & FAR (\%) $\downarrow$ \\
\hline
\multirow{2}{*}{Parakeet-TDT} & PCA & 6.87 & 13.81 & 0.81 \\
& Random & 50.08 $\pm$ 0.06 & 99.32 $\pm$ 0.24 & 0.69 $\pm$ 0.17 \\
\hline
\multirow{2}{*}{Moonshine} & PCA & 10.80 & 22.12 & 0.86 \\
& Random & 42.03 $\pm$ 1.14 & 97.46 $\pm$ 0.37 & 0.80 $\pm$ 0.09 \\
\hline
\end{tabular}
\vspace{-12pt}
\end{table}

To assess the importance of PCA-based subspace selection, we compared the PCA-based method with a random baseline at the same pruning locations. For the random baseline, the PCA transformation was replaced with a random permutation transformation. The first 50\% of the transformed channel dimensions were then retained. The results are summarized in Table~\ref{tab:pca_vs_random}. 

Random permutation without channel reduction yielded identical performance to the original model, confirming that the permutation itself does not affect detection performance. At 50\% channel reduction, however, the random baseline substantially increased EER and FRR for both Parakeet and Moonshine, while FAR remained comparable. These results indicate that, even at the same reduced dimensionality, the choice of retained dimensions is important, and that PCA-based selection preserves information relevant to WuW detection more effectively than random truncation.

\subsection{Calibration Source Comparison}

\begin{table}[t]
\centering
\caption{Effect of LibriSpeech (LS) and GSC v2 as calibration sources.}
\vspace{-10pt}

\label{tab:calibration_source}
\small
\setlength{\tabcolsep}{0.8pt}
\renewcommand{\arraystretch}{1.2}

\begin{subtable}{\linewidth}
\centering
\caption{Parakeet-TDT}
\label{tab:calibration_source_a}

\begin{tabular}{c|c|c|ccc}
\hline
\multirow{2}{*}[-0.5ex]{\makecell{Channel\\reduction}} &
\multirow{2}{*}[-0.5ex]{\makecell{Calibration\\data}} &
ASR &
\multicolumn{3}{c}{WuW} \\
\cline{3-6}
& &
WER(\%) $\downarrow$ &
EER(\%) $\downarrow$ &
FRR(\%) $\downarrow$ &
FAR(\%) $\downarrow$ \\
\hline

0 \%
& \makecell{LS / GSC v2}
& 2.76
& 6.67
& 11.84
& \textbf{0.79} \\
\hline

\multirow{2}{*}{50 \%}
& LS
& -
& \textbf{4.32}
& \textbf{9.84}
& \textbf{0.79} \\

& GSC v2
& -
& 6.87
& 13.81
& 0.81 \\

\hline
\end{tabular}
\end{subtable}

\par\medskip

\begin{subtable}{\linewidth}
\centering
\caption{Moonshine-base}
\label{tab:calibration_source_b}

\begin{tabular}{c|c|c|ccc}
\hline
\multirow{2}{*}[-0.5ex]{\makecell{Channel\\reduction}} &
\multirow{2}{*}[-0.5ex]{\makecell{Calibration\\data}} &
ASR &
\multicolumn{3}{c}{WuW} \\
\cline{3-6}
& &
WER(\%) $\downarrow$ &
EER(\%) $\downarrow$ &
FRR(\%) $\downarrow$ &
FAR(\%) $\downarrow$ \\
\hline

0 \%
& \makecell{LS / GSC v2}
& 5.73
& 9.67
& \textbf{18.81}
& 0.98 \\
\hline

\multirow{2}{*}{50 \%}
& LS
& -
& \textbf{9.66}
& 22.56
& 1.02 \\

& GSC v2
& -
& 10.80
& 22.12
& \textbf{0.86} \\

\hline
\end{tabular}
\end{subtable}

\end{table}

We approximately matched the total duration of the calibration data to examine the effect of the calibration source on WuW performance. As a comparison source, we selected LibriSpeech, a widely used benchmark corpus for ASR. Since the per-keyword calibration sets in GSC v2 span approximately 494.63--506.48 s, we sampled 501.5 s of speech from LibriSpeech train-clean-100 to construct a separate calibration set. Table~\ref{tab:calibration_source} reports the results at 50\% encoder channel reduction using each calibration source. WER is averaged over LibriSpeech test-clean and test-other, and WuW performance is evaluated on the GSC v2 test set.

\begin{figure}[t]
    \centering
    \begin{subfigure}[b]{0.48\columnwidth}
        \centering
        \includegraphics[width=\linewidth]{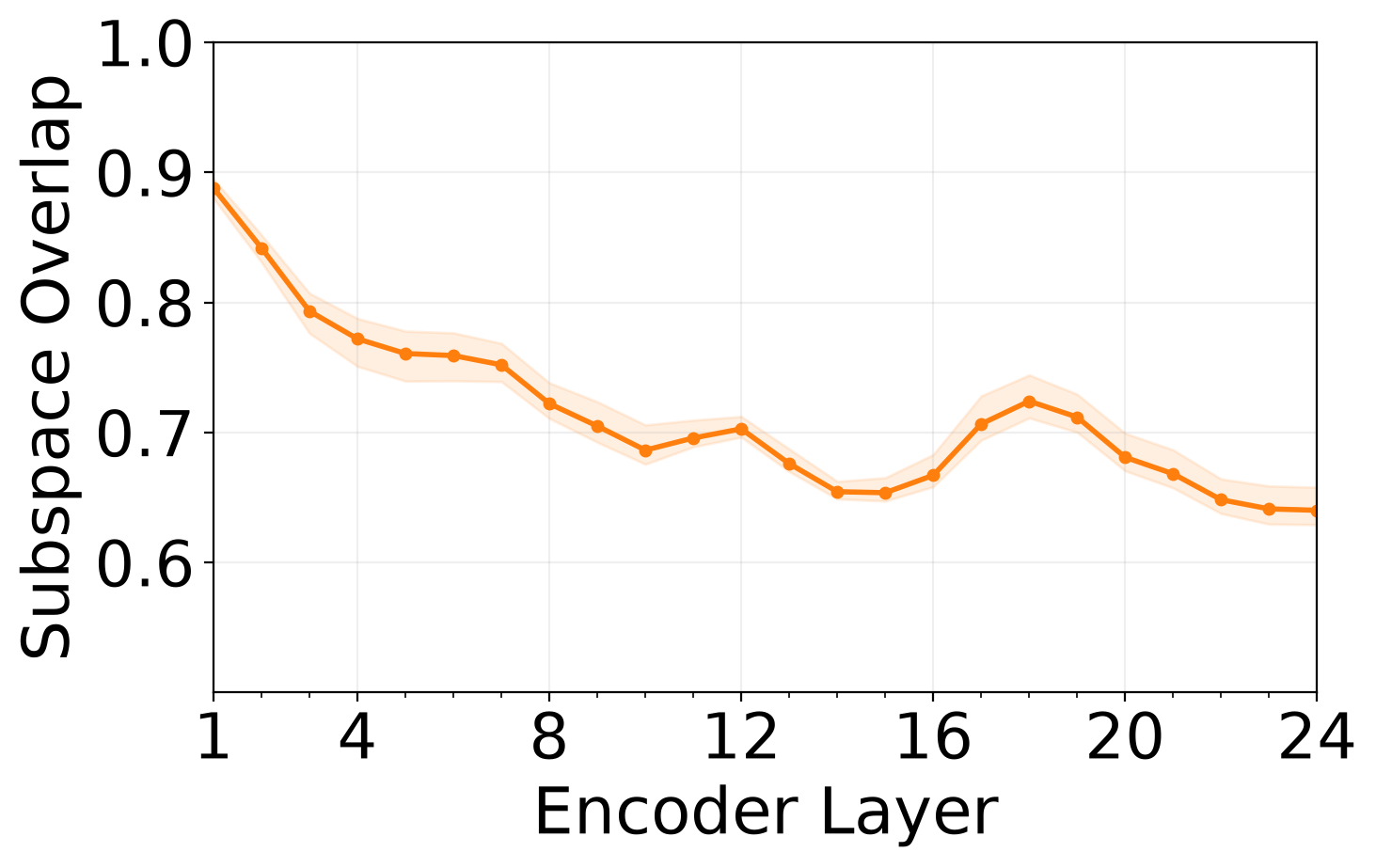}
        \caption{Parakeet-TDT}
        \label{fig:exp1_parakeet}
    \end{subfigure}
    \hfill
    \begin{subfigure}[b]{0.48\columnwidth}
        \centering
        \includegraphics[width=\linewidth]{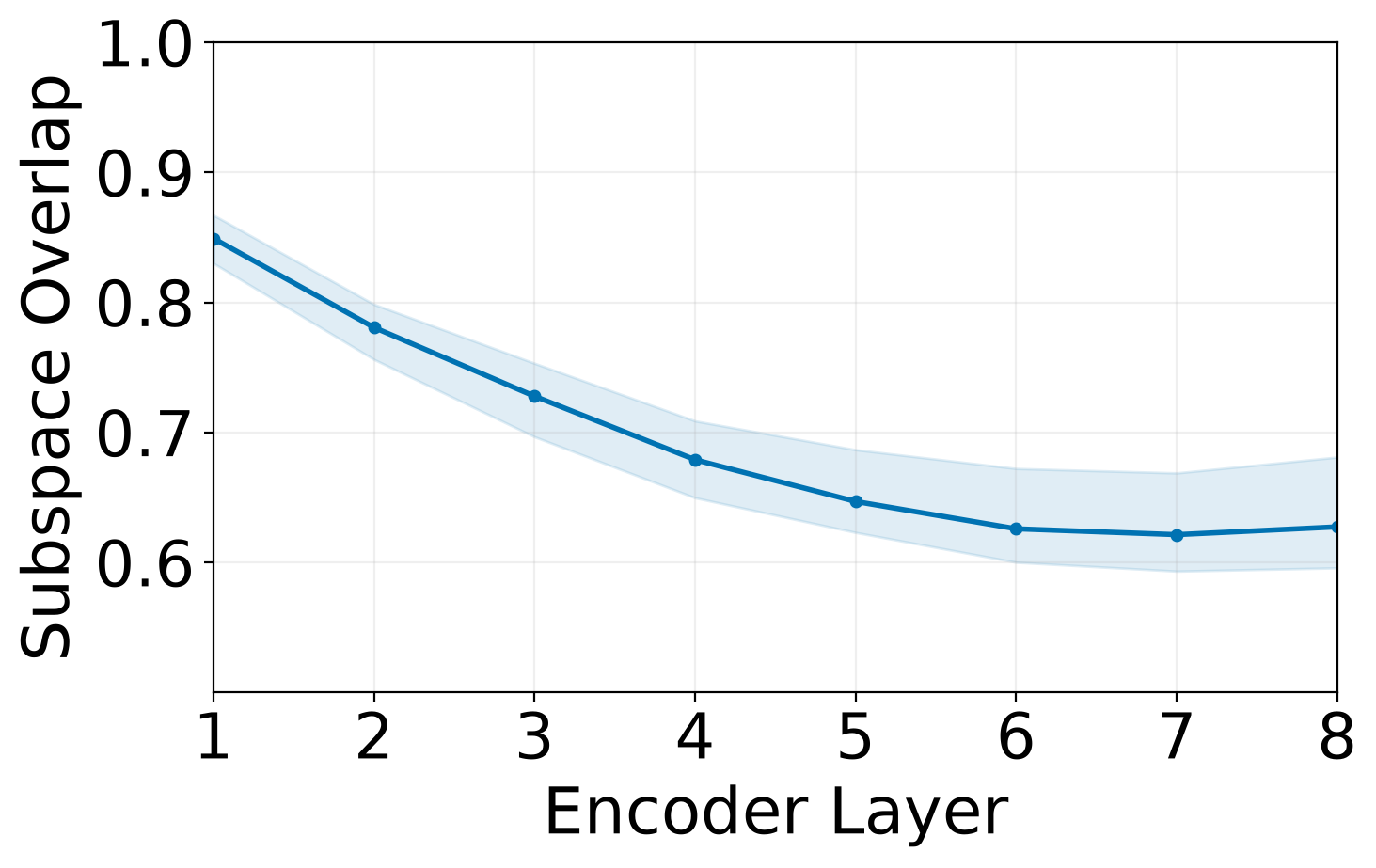}
        \caption{Moonshine-base}
        \label{fig:exp1_moonshine}
    \end{subfigure}
    \caption{Subspace overlap between PCA subspaces derived from LibriSpeech and GSC v2 across encoder layers at 50\% pruning. The solid line shows the mean across GSC v2 keywords, and the shaded region indicates the min–max range.}
    \vspace{-12pt}
    \label{fig:subspace}
\end{figure}

Consistent with Fig.~\ref{fig:pruning}(a), the 0\% pruning results confirm that the PCA-based transformation preserves the original ASR and WuW performance, with identical results across calibration sources. The original encoder results are therefore omitted from the table to avoid redundancy. Preservation of ASR capability after transformation is consistent with the goal of elastic speech inference, in which the original ASR function is retained while task-relevant compact encoders are derived.

For both models, the changes in WuW performance at 50\% channel reduction remained modest. The LibriSpeech calibration yielded a lower EER than GSC v2 for both, although the trend was not consistent across all WuW metrics. FAR remained broadly similar to the corresponding 0\% values. These results suggest that the PCA subspace estimated from the calibration data depends on the calibration source, and that its effect can vary across models and evaluation metrics.

The difference between the subspaces obtained from the two calibration sources is further examined in Fig~\ref{fig:subspace}. Subspace overlap is measured as the mean squared cosine of the principal angles between the PCA subspaces estimated from LibriSpeech and GSC v2. For both Parakeet and Moonshine, the two subspaces show relatively high overlap in the earlier layers, followed by an overall decrease as the encoder depth increases. The decreasing overlap may reflect differences in linguistic content between the calibration sources. This trend aligns with prior findings that higher speech-encoder layers capture more contextual and linguistic information \cite{he2025layerwise,pasad2021layerwise,shim2022selfattention}.

\subsection{Streaming Evaluation}
\begin{table}[t]
    \centering
    \caption{Streaming performance of Moonshine-base.}
    \label{tab:streaming_results}
    \setlength{\tabcolsep}{2.8pt}
    \renewcommand{\arraystretch}{1.0}

    \begin{tabular}{ccccc}
        \toprule
        Configuration &  \makecell{Encoder \\ Params(M)} & FA/h $\downarrow$ & FRR(\%) $\downarrow$ & RTF $\downarrow$ \\
        \midrule
        Original          & 20.15 & 0.47 & \textbf{4.51} & 0.2134 \\
        PCA-50\%       & 12.66 & \textbf{0.26} & 6.72 & \textbf{0.1718} \\
        \bottomrule
    \end{tabular}
    \vspace{-12pt}

\end{table}

Finally, we evaluate the PCA-based compression in a streaming setting on Hey-Snips. Given its smaller model scale, Moonshine-base is used for the streaming evaluation. Based on the calibration-source analysis, LibriSpeech is used for calibration, as it yielded lower EER for Moonshine-base. As shown in Table~\ref{tab:streaming_results}, 50\% channel reduction substantially decreases both the encoder parameter count and RTF. The compressed encoder yields a test FA/h of 0.26, remaining below the validation target of 0.5, while exhibiting a higher FRR. Meanwhile, it shows a lower RTF, indicating that the offline-selected compression setting also lowers inference cost.

%% file: Sections/5_conclusion.tex
\section{Conclusion}
We investigated training-free wake-up-word detection using pretrained ASR encoders and PCA-based structured pruning for compact encoder extraction. For both Moonshine-base and Parakeet-TDT, 50\% channel reduction largely retained WuW performance. At 0\% channel reduction, the transformed encoders preserved the original ASR WER regardless of the calibration source, confirming that the PCA-based transformation itself does not introduce degradation. 
These results provide initial evidence toward elastic speech inference, where task-relevant representation capacity can be selectively retained from a shared pretrained speech backbone. Future work will extend this approach to other speech front-end tasks and explore more effective task-relevant subspace extraction methods.

%% file: Sections/6_acknowledgement.tex
\section{ACKNOWLEDGMENT}
OpenAI ChatGPT was used to generate Fig.~\ref{fig:intro} from an author-provided concept and for language and grammar editing; all outputs were reviewed by the authors. 

This research was supported by Culture, Sports and Tourism R\&D Program through the Korea Creative Content Agency grant funded by the Ministry of Culture, Sports and Tourism in 2026 (Project Name: Development of AI-based personalized cultural and arts learning services using smart device, Project Number: RS-2026-25524629).

\section{COMPLIANCE WITH ETHICAL STANDARDS}
This study used only existing research datasets in accordance with their respective licenses. No additional ethical approval was required.
